\documentclass[10pt]{article}
\usepackage{graphicx}
\usepackage{amsmath,amssymb}
\newcommand{\rmd}{\mathrm{d}}
\newcommand{\rmi}{\mathrm{i}}
\newcommand{\bi}[1]{\textbf{\textit{#1}}}

\begin{document}
\title{\textbf{Excitons in single-walled carbon nanotubes: environmental
effect}}

\author{Oleksii A. Smyrnov\footnote{E-mails: smyrnov@onu.edu.ua, smyrnov.oleksii@gmail.com}\\
\emph{Department of Theoretical Physics, Odessa I. I. Mechnikov
National University,}\\ \emph{2 Dvoryanskaya St., Odessa 65026,
Ukraine}} \date{} \maketitle
\begin{abstract}
The properties of excitons in semiconducting single-walled carbon
nanotubes (SWCNTs), isolated in vacuum or a medium, and their
contributions to the optical spectra of nanotubes are studied
within the elementary potential model, in which an exciton is
represented as a bound state of two oppositely charged
quasi-particles confined to the nanotube surface. The emphasis is
given on the influence of dielectric environment surrounding a
nanotube on the exciton spectra. For nanotubes in environment with
permittivity less than $\sim1.8$ the ground-state exciton binding
energies exceed the respective energy gaps, while the obtained
binding energies of excitons in nanotubes in a medium with
permittivity greater than $\sim4$ are in good accordance with the
corresponding experimental data and consistent with known scaling
relation for the environmental effect. The stabilization process
of single-electron spectrum in SWCNTs in media with rather low
permittivities is discussed.
\end{abstract}

\section{Introduction}
\setcounter{equation}{0}

The most of experimental works on optical properties of
single-walled carbon nanotubes (SWCNTs) indicated that the exciton
contributions were dominant in the nanotubes optical
spectra~\cite{fluor}-\cite{bachilo2} and the exciton binding
energies were comparable to the corresponding energy
gaps~\cite{maul},\cite{wang}. These effects were explained
(predicted) in the theoretical works~\cite{ando}-\cite{jdd} on the
quasione-dimensional Wannier-like excitons in SWCNTs, which also
yielded large, in comparison with the 3D case, exciton binding
energies and revealed the determining influence of strong
interparticle Coulomb interaction in one dimension on the optical
properties of nanotubes. It was also shown (for example,
in~\cite{pereb}), that the potential of interaction between
electron and hole, which form an exciton, should be substantially
weakened by the dielectric environment surrounding a nanotube. The
dependence of exciton binding energy $\mathcal{E}$ on the
environmental permittivity $\varepsilon_\mathrm{env}$ obtained
in~\cite{pereb}:
$\mathcal{E}\sim\varepsilon^{-\alpha}_\mathrm{env}$ with
$\alpha=1.4$ is not similar to that in the 3D case ($\alpha=2$).
As it was pointed out in~\cite{pereb}, the relation
$\mathcal{E}\sim\varepsilon^{-\alpha}_\mathrm{env}$ with
$\alpha=1.4$ is accurate only for nanotubes surrounded by media
with high permittivity ($\varepsilon_\mathrm{env}\gtrsim4$),
because the results of calculations based on the Ohno potential
chosen in~\cite{pereb} to model the unscreened Coulomb interaction
between carbon $\pi$-orbitals is not sensitive to the value of the
potential parameter only if the exciton radius is rather large.
This takes place in the environmental permittivities range, which
is of technological interest (silicon oxide environment,
etc.~\cite{pereb}). However, in the most of experiments on optical
response of SWCNTs individual nanotubes were isolated in media
with low permittivities: the hydrocarbon environment of SDS
micelles~\cite{fluor}-\cite{maul},~\cite{wang2} (by~\cite{kiow}
the corresponding empirical $\varepsilon_\mathrm{env}=2-2.5$,
while~\cite{maul} used $\varepsilon_\mathrm{env}\sim4$); the
polymer matrix environment~\cite{wang} (used
$\varepsilon_\mathrm{env}=2.5$); air~\cite{kiow},\cite{ohno}
($\varepsilon_\mathrm{env}=1$). This is why here, using an exciton
model, which is not influenced by the exciton radius but depends
on the tube radius and carriers effective masses (and thus on the
nanotube chirality), we apply the scaling relation for exciton
binding energies from~\cite{pereb} to excitons in SWCNTs in
low-permittivity media ($1\leq\varepsilon_\mathrm{env}\lesssim4$)
to obtain the corresponding scaling parameter $\alpha$.

The exciton in a SWCNT is modelled here as a bound state of two
quasi-particles, which opposite charges are smeared uniformly
along infinitesimal narrow bands at the tube surface, with the
interaction potential having the Coulomb attraction tail (see
section~2, or~\cite{adsmyr},\cite{jpcs} for details). The
single-electron spectrum and wave functions were obtained like
in~\cite{adtish},\cite{tish} by the zero-range potentials
method~\cite{aghh},\cite{demk}. It turned out, that within the
mentioned model the binding energies of excitons in the ground
state in nanotubes surrounded by a medium with
$\varepsilon_\mathrm{env}\sim4$ were in good accordance with the
corresponding experimental data from~\cite{maul} and in the range
$\varepsilon_\mathrm{env}\in[4,~16]$ obey the scaling relation
from~\cite{pereb} with $\alpha\simeq1.4$. Moreover, for the same
$\varepsilon_\mathrm{env}\sim4$ the differences between the
ground-state exciton binding energies and those of the lowest
excited states are also in good agreement with the respective
experimental results of~\cite{maul} and~\cite{wang} (section~3).
In the region $\varepsilon_\mathrm{env}\lesssim1.8$ the
ground-state exciton binding energies exceed the corresponding
energy gaps. This leads to instability of the nanotube
single-electron states with regard to formation of exctions,
which, however, become stabilized because of the additional
screening effect stipulated by born excitons. Since some of the
single-electron states have transformed into excitons the edges of
the forbidden band move apart, and this results in the enhancement
(blueshift) of the lowest optical transition energy like in
experiments~\cite{kiow},~\cite{ohno}. The corresponding estimates
are in satisfactory agreement with results of~\cite{kiow}.
Besides, in the ranges of low environmental permittivities
$\varepsilon_\mathrm{env}\in[1,~1.75]$ and
$\varepsilon_\mathrm{env}\in[2,~4.5]$ the ground-state exciton
binding energies satisfy the mentioned relation from~\cite{pereb}
with slightly smaller values of parameter $\alpha$: 1.121 and
1.258, respectively (see section~3).

\section{Model of exciton in a semiconducting SWCNT}
\setcounter{equation}{0}

By analogy with the 3D case it can be shown (like
in~\cite{adsmyr}) that within the long-wave approximation the wave
equation for the Fourier transform $\phi$ of envelope function in
the wave packet from products of the electron and hole Bloch
functions, which represents a two-particle state of large-radius
rest exciton in a quasione-dimensional semiconducting nanotube
with the longitudinal period $a$, is reduced to the following 1D
Schr\"{o}dinger equation:
\begin{equation}\label{2.1}
\begin{split}
-\frac{\hbar^2}{2\mu}\phi''(z)+V(z)&\phi(z)=\mathcal{E}\phi(z),\
\mathcal{E}=E_\mathrm{exc}-E_\mathrm{g},\\
&-\infty<z<\infty,
\end{split}
\end{equation}
with the exciton reduced effective mass $\mu$, the forbidden band
width $E_\mathrm{g}$ and the electron-hole (e-h) interaction
potential
\begin{equation}\label{2.2}
\begin{split}
V(z)=-\int\limits_{\mathrm{E}_3^a}\int\limits_{\mathrm{E}_3^a}&\frac{e^2|u_{\mathrm{c};0}(\bi{r}_1)|^2|u_{\mathrm{v};0}(\bi{r}_2)|^2\rmd\bi{r}_1\rmd\bi{r}_2}{\left((x_1-x_2)^2+(y_1-y_2)^2+(z+z_1-z_2)^2 \right)^{1/2}},\\
&\mathrm{E}_3^a=\mathrm{E}_2\times(0<z<a),
\end{split}
\end{equation}
where $u_{\mathrm{c,v};k}(\bi{r})$ are the Bloch amplitudes of the
Bloch wave functions $\psi_{\mathrm{c,v};k}(\bi{r})=\exp(\rmi
kz)u_{\mathrm{c,v};k}(\bi{r})$ of the conduction and valence band
electrons of a SWCNT, respectively, and $k$ is the electron
quasi-momentum. Assuming that the charges of electron and hole,
which participate in the formation of exciton, are smeared
uniformly along infinitesimal narrow bands at the nanotube surface
we obtain from~(\ref{2.2}):
\begin{equation}\label{2.3}
V_{R_0}(z)=-\frac{2e^2}{\pi|z|}\mathrm{K}\left[-\frac{4R^2_0}{z^2}\right],
\end{equation}
where $\mathrm{K}$ is the complete elliptic integral of the first
kind and $R_0$ is the nanotube radius. This potential is the
simplest approximation to the bare Coulomb potential, which
accounts the finiteness of the nanotube diameter. At this point of
the model SWCNTs differ only by their radii and the carriers
effective masses. However, even the combination of these
parameters allows to specify the nanotube chirality.

To take into account the screening of e-h interaction potential by
the nanotube band electrons we have applied the Lindhard method
(the so-called random phase approximation), according to which the
analogue of quasione-dimensional electrostatic
potential~(\ref{2.3}) screened by the nanotube $\pi$-electrons is
given by the following expression~\cite{adsmyr}:
\begin{equation}\label{2.4}
\varphi_{R_0}(z)=-\frac{2e^2}{\pi
R_0}\int\limits^{\infty}_0\frac{I_0(q)K_0(q)\cos(qz/R_0)}{1+g_aq^2I_0(q)K_0(q)}\
\rmd q,
\end{equation}
where $I$ and $K$ are the modified Bessel functions of the first
and second kind, respectively, and the dimensionless screening
parameter
\begin{equation}\label{2.5}
g_a=\frac{e^2\hbar^4}{\pi m^2_\mathrm{b}
R^2_0}\sum\limits_s\int\limits^{\pi/a}_{-\pi/a}\frac{1}{E^3_{\mathrm{g};s,s}(k)}
\left|\left\langle\psi_{\mathrm{c};k,s}\left|\frac{\partial}{\partial
z}\right|\psi_{\mathrm{v};k,s}\right\rangle\right|^2\rmd k.
\end{equation}
Here $s$ numbers $\pi$-electron bands, which are mirror with
respect to the Fermi level, because only for those bands the
matrix element in~(\ref{2.5}) is
nonzero~\cite{adsmyr},\cite{adtish}, and
$m_\mathrm{b}=0.415m_\mathrm{e}$ is the bare mass
from~\cite{adtish},\cite{tish}.

As it was mentioned above, most of existing experiments on the
optical response of SWCNTs~\cite{fluor}-\cite{wang} dealt with
nanotubes isolated not in vacuum but in media with the dielectric
constants different from unity, therefore the corresponding
screening of the e-h interaction potential should be also taken
into account because a dielectric medium, surrounding a nanotube,
should noticeably change the e-h interaction potential. For
example, in experimental
works~\cite{bachilo}-\cite{maul},\cite{wang2}, which used the
methods described in~\cite{fluor}, investigated isolated SWCNTs
were encased in the sodium dodecyl sulfate (SDS) cylindrical
micelles disposed in heavy water. Because of these SDS micelles,
which provided a pure hydrocarbon environment around individual
nanotubes, the high permittivity solvent $\mathrm{D_2O}$ did not
reach nanotubes. However, the environment of hydrophobic
hydrocarbon "tails" $(-\mathrm{C_{12}H_{25}})$ of the SDS
molecules has the permittivity greater than unity (by the
experiment~\cite{kiow} it is about $2-2.5$). In accordance with
figure~1A from~\cite{fluor} and with~\cite{banyai} a simple model
of a SWCNT in a dielectric environment was considered, namely: a
narrow, infinite cylinder with radius $R_0$ in a medium with the
dielectric constant $\varepsilon_\mathrm{env}$ and some internal
screening parameter $\varepsilon_\mathrm{int}$. The corresponding
analogue of potential~(\ref{2.3}) screened by environment within
the framework of the mentioned model is given by~\cite{jpcs}:
\begin{equation}\label{2.6}
\varphi_{R_0}(z)=-\frac{2e^2}{\pi
R_0}\int\limits^\infty_0\frac{I_0(q)K_0(q)\cos(qz/R_0)}{[\varepsilon_\mathrm{env}
I_0(q)K_1(q)+\varepsilon_\mathrm{int}I_1(q)K_0(q)]q}\rmd q.
\end{equation}
The internal screening parameter
$\varepsilon_\mathrm{int}\equiv\varepsilon_\mathrm{int}(q)=1+g_aq^2I_0(q)K_0(q)$
according to~(\ref{2.4}). As it will be shown further (section~3),
potential~(\ref{2.6}) with this $\varepsilon_\mathrm{int}$ can be
used to model the e-h interaction in SWCNTs isolated in medium
with $\varepsilon_\mathrm{env}\gtrsim1.8$ (e.g.: the SDS
environment~\cite{maul}, the polymer matrix~\cite{wang}, etc.).

\section{Calculation results. Environmental screening influence}
\setcounter{equation}{0}

The exciton binding energies and envelope functions were obtained
within the above-stated exciton model using the wave
equation~(\ref{2.1}) with the different e-h interaction
potentials~(\ref{2.4}),~(\ref{2.6}) and the single-electron
parameters (effective masses, single-particle wave functions, band
gaps) calculated according~\cite{adtish},\cite{tish} within the
zero-range potentials method~\cite{aghh},\cite{demk}.

According to the wave equation~(\ref{2.1}) with the e-h
interaction potential~(\ref{2.4}) screened only by the nanotube
band electrons and that screened also by the external dielectric
medium~(\ref{2.6}) we have calculated the binding energies of
excitons in different SWCNTs in vacuum and in the SDS environment,
respectively (see table 1).\footnote{The ground state of exciton
corresponds to the even envelope function $\phi(z)$ ($z$ is the
distance along the tube axis between electron and hole) and the
lowest excited state corresponds to the odd one, further the
excited states of different parity actually alternate.} The
experimental value of dielectric constant of the SDS environment
$\varepsilon_\mathrm{env}=2\div2.5$ was taken from~\cite{kiow}.
Table 1 shows that for these values of $\varepsilon_\mathrm{env}$
there is only a qualitative similarity of the obtained results to
the corresponding data from experimental work~\cite{maul}, though
this $\varepsilon_\mathrm{env}$ is taken from experiment. However,
if we, following~\cite{maul}, choose
$\varepsilon_\mathrm{env}=4.4$ the ground-state exciton binding
energies become almost identical to those obtained in~\cite{maul}
(see table 2). Moreover, for each considered SWCNT the obtained
difference between the ground-state exciton binding energy and
that of exciton in the lowest exited state also becomes almost
equal to the respective experimental value
from~\cite{maul}.\footnote{Recall, that the difference between the
binding energies of exciton in two different states and the
difference between the corresponding excitation energies are
equal, so one can compare
$\mathcal{E}_{0;\mathrm{even}}-\mathcal{E}_{1;\mathrm{odd}}$ and
$E^\mathrm{2g}_{11}-E^\mathrm{1u}_{11}$ from~\cite{maul} and
$E_\mathrm{2p}-E_\mathrm{1s}$ from~\cite{wang}. Recall also, that
the exciton states in~\cite{maul} with the even $z$-inversion
symmetry were indexed by 1 and those with the odd one by 2.} The
discrepancies between the data from table 2 and the corresponding
results from~\cite{maul} for the exciton binding energies and also
for the differences
$\mathcal{E}_{0;\mathrm{even}}-\mathcal{E}_{1;\mathrm{odd}}$ and
$E^\mathrm{2g}_{11}-E^\mathrm{1u}_{11}$ appears to be inessential
if the variation of $\varepsilon_\mathrm{env}$ in $\pm0.3$ for
different tubes in~\cite{maul} is taken into account. It is also
worth mentioning, that the obtained here differences
$\mathcal{E}_{0;\mathrm{even}}-\mathcal{E}_{1;\mathrm{odd}}$ are
also very close to the respective results of
experiment~\cite{wang} on SWCNTs isolated in polymer matrixes.

Table 1 also shows that the ground-state binding energies of
excitons in nanotubes in vacuum are substantially larger than the
corresponding band gaps, while those in nanotubes in the medium
with $\varepsilon_\mathrm{env}\gtrsim2$ occur already inside of
the respective band gaps. The applicability of effective-mass
approximation may seem questionable for such large exciton binding
energies $\mathcal{E}_{0;\mathrm{even}}$ in vacuum. However, it is
not the effective-mass approximation that causes so great absolute
value of the exciton ground-state energy. The matter is that the
effective-mass approximation consists in the replacement of the
original dispersion relations for the valence and conduction
bands, which come in the equation for exciton envelope function as
$\epsilon_\mathrm{c}(k)-\epsilon_\mathrm{v}(-k)$, by their
expansion to the quadratic terms
$\epsilon_\mathrm{c}(k)-\epsilon_\mathrm{v}(-k)\simeq
E_\mathrm{g}+\hbar^2k^2/2\mu$. But such a replacement only
increases the kinetic part of exciton energy operator and thus can
only reduce the absolute values of (negative) exciton binding
energies. Therefore, without the effective-mass approximation, the
exciton binding energies come even larger. Besides, the
calculations of the exciton radii (for example, as the
root-mean-square deviation of the envelope function Fourier
transform $\phi(z)$ from the origin on the tube axis) show that
for the ground state in vacuum they are of the order of nanotube
diameter $2R_0$, which is much larger than the nanotube
longitudinal period $a$, which in its turn is of the order of tube
lattice parameter ($\sim0.142$~nm for CNTs). Thus, the long-wave
approximation formalism is also applicable. Later on we will
return to the discussion of seeming instability of the nanotube
single-electron states with respect to formation of excitons in
SWCNTs in media with $\varepsilon_\mathrm{env}<2$, and for now let
us consider in greater detail the interval of environmental
dielectric constants $\varepsilon_\mathrm{env}\gtrsim2$.

\begin{table}[t]
\begin{center}
\caption{The ground-state exciton binding energies
$\mathcal{E}_\mathrm{0;even}$ for different SWCNTs in vacuum
(according to the wave equation~(\ref{2.1}) with screened
potential~(\ref{2.4})) and in the medium with
$\varepsilon_\mathrm{env}=2\div2.5$ from~\cite{kiow} (according
to~(\ref{2.1}) with screened potential~(\ref{2.6})), and also the
corresponding results from experimental work~\cite{maul}.}
\begin{tabular}{|c|c|c|c|c|c|c|}
\hline
$\mathrm{Chirality}$&$2R_0~(\mathrm{nm})$&$\mu~(m_\mathrm{e})$&$E_\mathrm{g}~(\mathrm{eV})$&$\mathcal{E}_{0;\mathrm{even}}~(\mathrm{eV})$&$\mathcal{E}_{0;\mathrm{even}}~(\mathrm{eV})$&$E^\mathrm{b}_{11}~(\mathrm{eV})$\\
&&&&$\mathrm{in~vacuum}$&$\mathrm{in~medium}$&\cite{maul}\\
&&&&&$\varepsilon_\mathrm{env}=2\div2.5$&\\
\hline
(6, 4)&0.6825&0.0651&1.21&2.53&$1.09\div0.82$&0.38\\
\hline
(6, 5)&0.7468&0.0510&1.10&2.25&$0.95\div0.71$&0.33\\
\hline
(9, 1)&0.7468&0.0748&1.117&2.46&$1.07\div0.81$&0.38\\
\hline
(8, 3)&0.7711&0.0644&1.076&2.32&$1.00\div0.75$&0.35\\
\hline
(7, 5)&0.8174&0.0530&1.01&2.10&$0.90\div0.68$&0.28\\
\hline
(9, 4)&0.9029&0.0522&0.9176&1.95&$0.84\div0.63$&0.33\\
\hline
\end{tabular}
\end{center}
\end{table}

\begin{table}[t]
\begin{center}
\caption{The ground-state exciton binding energies
$\mathcal{E}_\mathrm{0;even}$ for different SWCNTs in the medium
with $\varepsilon_\mathrm{env}=4.4$ from~\cite{maul} according
to~(\ref{2.1}) with screened potential~(\ref{2.6}), the difference
between the exciton binding energy in the ground state and first
excited one
$\mathcal{E}_{0;\mathrm{even}}-\mathcal{E}_{1;\mathrm{odd}}$, and
also the corresponding experimental data from~\cite{maul}
and~\cite{wang}.}
\begin{tabular}{|c|c|c||c|c|c|}
\hline
$\mathrm{Chirality}$&$\mathcal{E}_{0;\mathrm{even}}~(\mathrm{eV})$&$E^\mathrm{b}_{11}~(\mathrm{eV})$&$\mathcal{E}_{0;\mathrm{even}}-\mathcal{E}_{1;\mathrm{odd}}~(\mathrm{eV})$&$E^\mathrm{2g}_{11}-E^\mathrm{1u}_{11}~(\mathrm{eV})$&$E_\mathrm{2p}-E_\mathrm{1s}~(\mathrm{eV})$\\
&in medium&\cite{maul}&&\cite{maul}&\cite{wang}\\
&$\varepsilon_\mathrm{env}=4.4$&&&&\\
\hline
(6, 4)&0.39&0.38&0.346&0.325&-\\
\hline
(6, 5)&0.33&0.33&0.302&0.285&0.31\\
\hline
(9, 1)&0.38&0.38&0.340&0.315&-\\
\hline
(8, 3)&0.36&0.35&0.317&0.295&0.3\\
\hline
(7, 5)&0.32&0.28&0.287&0.24&0.28\\
\hline
(9, 4)&0.30&0.33&0.267&0.28&-\\
\hline
\end{tabular}
\end{center}
\end{table}

To reveal the general dependence of binding energies of excitons
in SWCNTs on the environmental dielectric constant the
corresponding scaling relation from~\cite{pereb} is applied.
According to~\cite{pereb} after taking the logarithm one can
obtain from the mentioned relation:
\begin{equation}\label{3.1}
\ln\frac{\mathcal{E}}{\mathcal{E}'}\approx(\alpha-2)\ln\frac{R_0}{R'_0}+(\alpha-1)\ln\frac{\mu}{\mu'}-\alpha\ln\frac{\varepsilon_\mathrm{env}}{\varepsilon'_\mathrm{env}},
\end{equation}
where $\alpha$ according to~\cite{pereb} equals 1.40, and
$\mathcal{E}',~R'_0,~\mu',~\varepsilon'_\mathrm{env}$ are some
magnitudes, which do not influence on the relation, and introduced
here just to make the corresponding variables dimensionless. The
ground-state exciton binding energies obtained using the wave
equation~(\ref{2.1}) with potential~(\ref{2.6}) were substituted
into~(\ref{3.1}) for the set of SWCNTs with different diameters
($2R_0\in[0.63,~2.19]$~nm) surrounded by media with permittivities
in the range of interest indicated by~\cite{maul} and~\cite{kiow}
($\varepsilon_\mathrm{env}\in[2,~4.5]$). Using the least-squares
method we found that for these ranges of the nanotubes diameters
and environmental permittivities relation~(\ref{3.1}) was valid
when $\alpha\simeq1.258$ (see figure 1(a)). It should be noted,
that for the same set of nanotubes, but for the environmental
dielectric constants range $\varepsilon_\mathrm{env}\in[4,~16]$
the calculated value of $\alpha$ directly approaches that obtained
in~\cite{pereb}. According to~\cite{pereb} the scaling relation
with $\alpha=1.40$ is accurate only in the region
$\varepsilon_\mathrm{env}\gtrsim4$, this explains the discrepancy
between $\alpha$ obtained here and that from~\cite{pereb} in the
region $\varepsilon_\mathrm{env}\lesssim4$.

The binding energies of excitons in the first excited state in
SWCNTs also obey relation~(\ref{3.1}), but with $\alpha\simeq1.89$
for the range $\varepsilon_\mathrm{env}\in[2,~4.5]$ (see figure
1(b)) and with $\alpha\simeq1.93$ for the range
$\varepsilon_\mathrm{env}\in[4,~16]$. This is clear because the
radii of excitons in SWCNTs even in the first excited state
(especially in media with large $\varepsilon_\mathrm{env}$) are
close to those of the 3D excitons, for which, as is well known,
$\alpha=2$.

\begin{figure}[t]
\begin{center}
\includegraphics[scale=0.65]{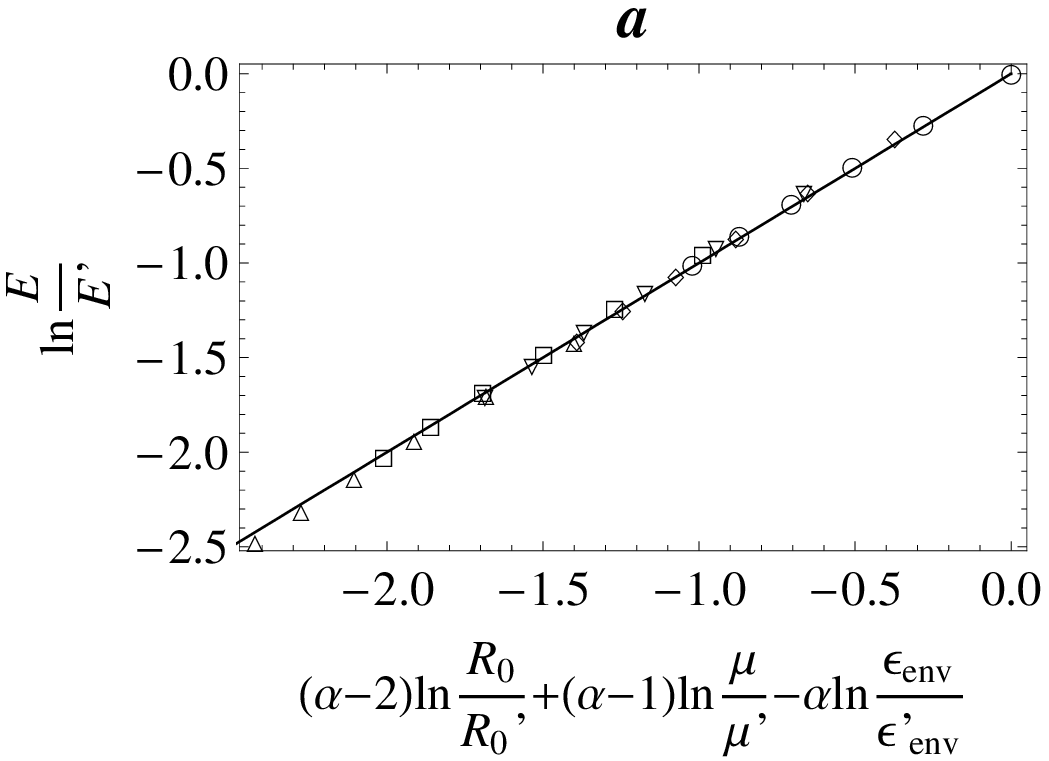}
\includegraphics[scale=0.65]{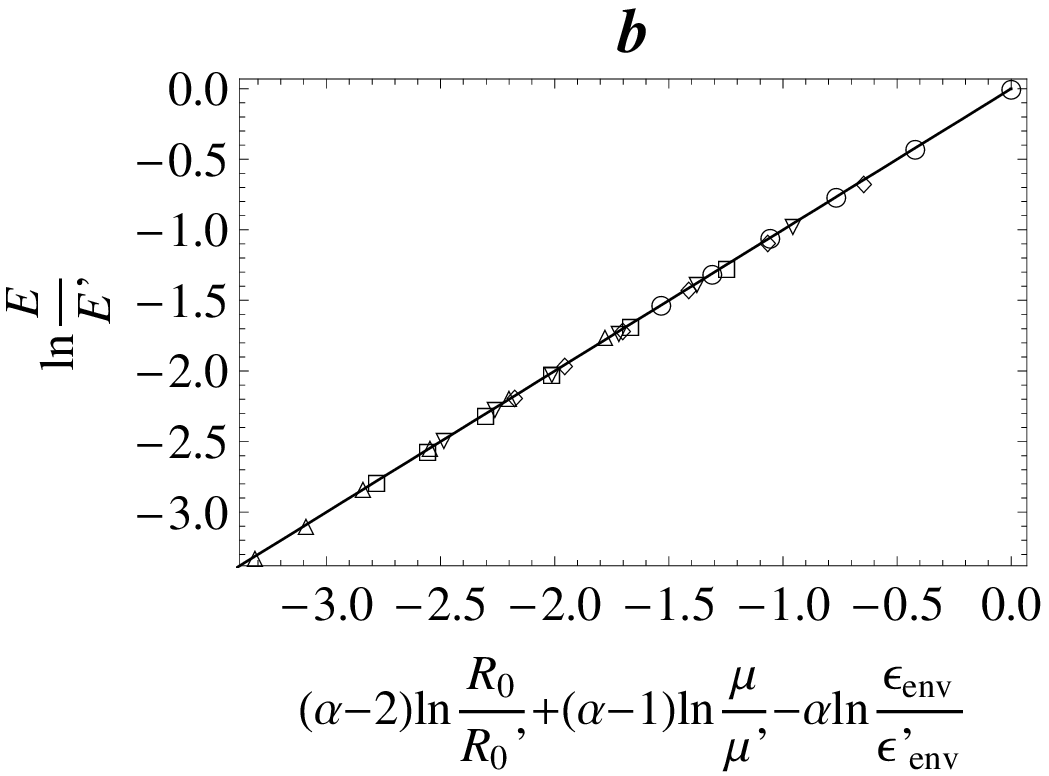}
\includegraphics[scale=0.65]{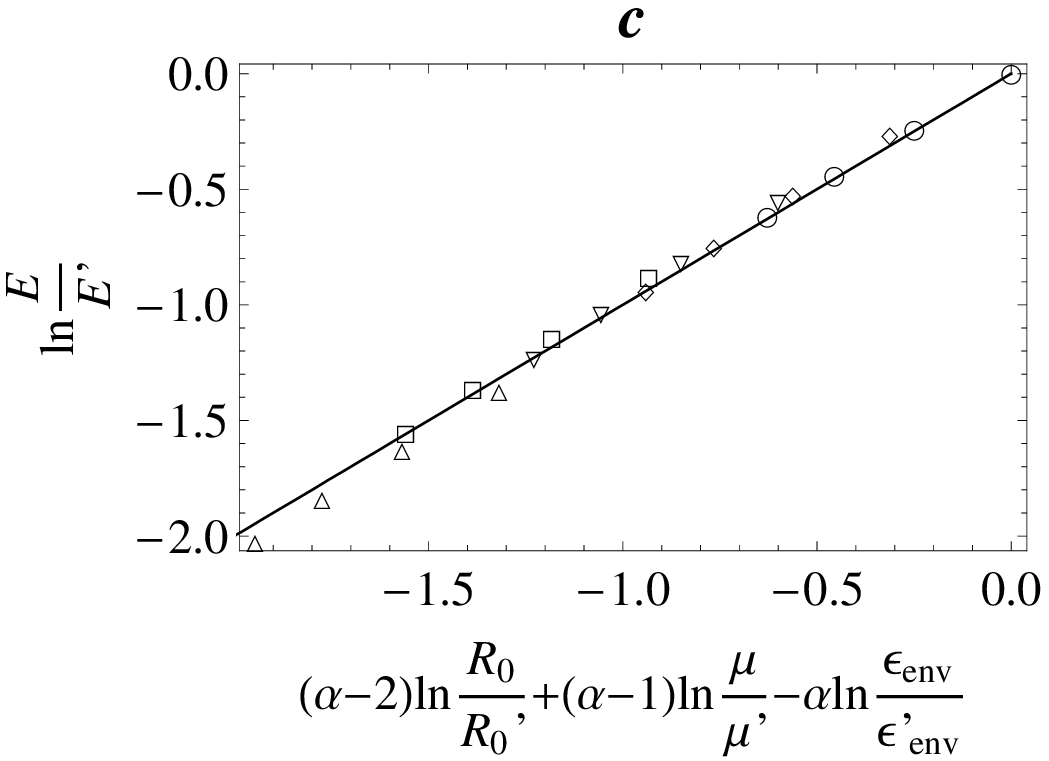}
\includegraphics[scale=0.65]{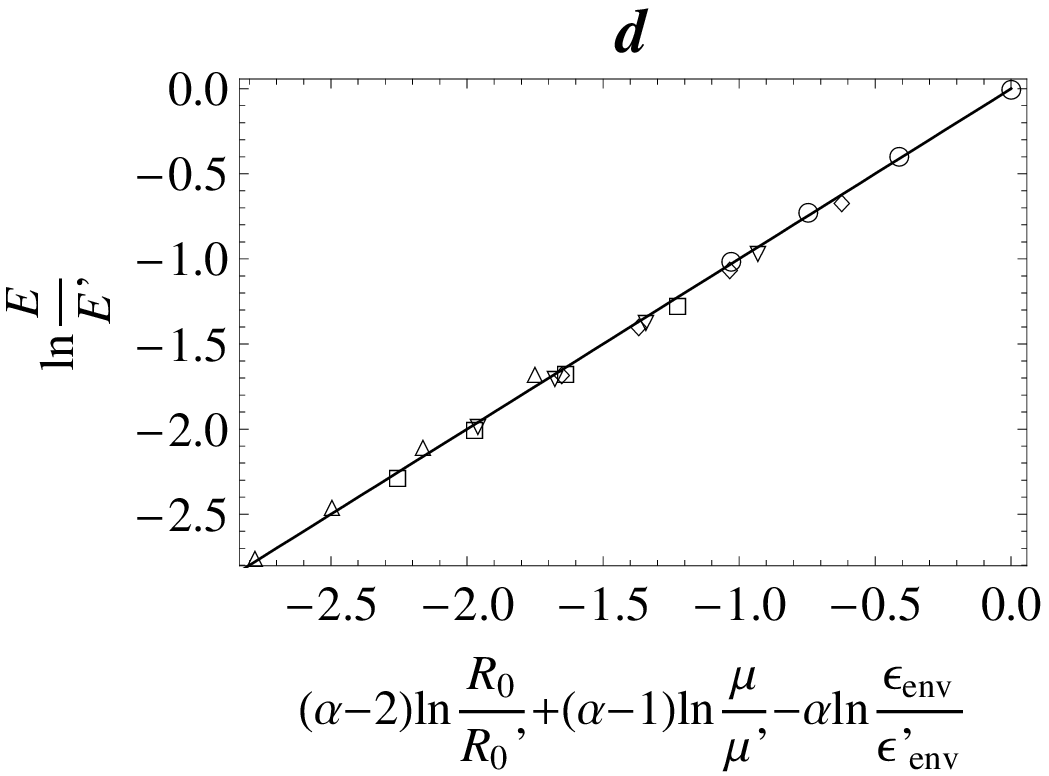}
\end{center}
\caption{General dependence of the binding energy of exciton in
SWCNTs of different chirality types ((8,~0)~$\circ$,
(7,~5)~$\diamond$, (9,~7)~$\triangledown$, (15,~7)~$\square$ and
(28,~0)~$\vartriangle$ with diameters: $\sim$~0.63, 0.82, 1.09,
1.52 and 2.19~nm, and band gaps: $\sim$~1.42, 1.01, 0.76, 0.54 and
0.37~eV, respectively) on the environmental dielectric constant
$\varepsilon_\mathrm{env}$, the nanotube radius $R_0$ and the
carriers effective masses $\mu$. Abscissa is the right side
of~(\ref{3.1}) and ordinate is the left one. Black solid lines are
plotted using the least-squares method. (a) the dependence of the
binding energies of excitons in the ground state in SWCNTs
surrounded by media with $\varepsilon_\mathrm{env}\in[2,~4.5]$
($\varepsilon_\mathrm{env}$ changes with step 0.5), the points
correspond to~(\ref{3.1}) with $\alpha=1.258$, the primed
magnitudes from~(\ref{3.1}) are the respective parameters of the
(8,~0) nanotube in medium with $\varepsilon_\mathrm{env}=2$; (b)
the points similar to (a), but for the first excited exciton state
and $\alpha=1.89$; (c) the dependence of the unstabilized binding
energies of excitons in the ground state in SWCNTs surrounded by
media with $\varepsilon_\mathrm{env}\in[1,~1.75]$
($\varepsilon_\mathrm{env}$ changes with step 0.25), the points
correspond to~(\ref{3.1}) with $\alpha=1.121$, the primed
magnitudes from~(\ref{3.1}) are the respective parameters of the
(8,~0) nanotube in vacuum; (d) the points similar to (c), but for
the first excited exciton state and $\alpha=1.838$.}
\end{figure}

As it was mentioned above, the obtained binding energies of
excitons in the ground state in semiconducting SWCNTs in vacuum
appeared to be larger than the corresponding band gaps (see table
1). More precisely, for the considered set of nanotubes surrounded
by medium with $\varepsilon_\mathrm{env}=1.75-1.85$ the
ground-state exciton binding energy becomes equal to the
corresponding energy gap. For smaller values of
$\varepsilon_\mathrm{env}$ the ground-state exciton binding energy
exceeds the respective energy gap, and this may lead to
instability of the nanotubes single-electron states with regard to
formation of excitons. However, the incipient excitons induce an
additional screening effect stipulated by their rather great
polarizability in the longitudinal electric field. This effect
essentially weakens the e-h interaction. Under certain critical
concentration of excitons the ground-state exciton binding energy
becomes smaller than the energy gap and the conversion of
single-electron states into excitons ends. The upper and lower
limits of the exciton concentration $n$ can be given as
follows~\cite{jpcs}:
\begin{equation}\label{3.2}
\frac{\varepsilon_\mathrm{exc}-1}{4\pi}\frac{\mathcal{E}_0-\mathcal{E}_1}{2e^2}\left|
\int\limits^\infty_{-\infty}z^2|\phi_0(z)|^2\rmd z\right|^{-1}\leq
n\leq\frac{\varepsilon_\mathrm{exc}-1}{4\pi}\frac{\mathcal{E}_0-\mathcal{E}_1}{2e^2}\left|
\int\limits^{\infty}_{-\infty}z\phi_0(z)\phi_1(z)\rmd
z\right|^{-2},
\end{equation}
where $\phi$ is the component of Fourier transform of the
corresponding exciton envelope function, it depends only on the
distance $z$ between the electron and hole along the tube axis.
Each $\phi$ is the solution of wave equation~(\ref{2.1}) with
potential~(\ref{2.6}), where $\varepsilon_\mathrm{env}\lesssim1.8$
and
$\varepsilon_\mathrm{int}\equiv\varepsilon_\mathrm{int}(q)=\varepsilon_\mathrm{exc}+1+g_aq^2I_0(q)K_0(q)$
(according to~(\ref{2.4}) and~(\ref{2.5})). The ground-state
envelope function $\phi_0$ is the even solution of the 1D
Schr\"{o}dinger equation~(\ref{2.1}), which satisfies the boundary
condition $\phi'(0)=0$, $\phi_1$ is the odd solution
of~(\ref{2.1}), which corresponds to the lowest excited exciton
state and satisfies the boundary condition $\phi(0)=0$,
$\mathcal{E}_0$ and $\mathcal{E}_1$ are the corresponding exciton
binding energies (eigenvalues of~(\ref{2.1})), and
$\varepsilon_\mathrm{exc}$ is the incipient excitons contribution
to the dielectric function of a nanotube. Varying
$\varepsilon_\mathrm{exc}$ in~(\ref{2.6}) substituted into wave
equation~(\ref{2.1}) one can match $\mathcal{E}_0$ to the energy
gap. Further, $\mathcal{E}_1$ can be obtained from the same
equation with the fixed $\varepsilon_\mathrm{exc}$ and with the
corresponding boundary condition. These magnitudes allow to
calculate from~(\ref{3.2}) the upper and lower limits for the
critical concentration of excitons $n_\mathrm{c}$. Using the
obtained $n_\mathrm{c}$ one can calculate the shift of the
forbidden band edges, which move apart due to the transformation
of some single-electron states into excitons. This results in the
enhancement of energy gap and hence within the effective mass
approximation the lowest optical transition energy $E_{11}$ should
be blueshifted by
\begin{equation}\label{3.3}
\delta E_{11}=\frac{(\pi\hbar\widetilde{n}_\mathrm{c})^2}{2\mu},
\end{equation}
like in experiments~\cite{kiow} and~\cite{ohno}. Here
$\widetilde{n}_\mathrm{c}=n_\mathrm{c}\pi R_0^2$ is the linear
critical concentration of excitons.

According to experiment~\cite{kiow} this blueshift is about
$40-55$~meV for SWCNTs in air (vacuum,
$\varepsilon_\mathrm{env}=1$) with respect to those encased in SDS
micelles~\cite{bachilo2} (in this case according to~\cite{kiow}
$\varepsilon_\mathrm{env}$, at least, larger than 2).
By~(\ref{3.3}) this blueshift gives the linear critical
concentration of excitons $\widetilde{n}_\mathrm{c}$, which should
be born in a SWCNT to stabilize the nanotube single-electron
spectrum, about $80~\mathrm{\mu m}^{-1}$ for nanotubes with
diameters $\sim1$~nm and about $50~\mathrm{\mu m}^{-1}$ for
nanotubes with diameters $\sim1.5-2$~nm. The corresponding
estimates in accordance with~(\ref{3.2}) are about
$100-150~\mathrm{\mu m}^{-1}$ for SWCNTs with diameters $\sim1$~nm
(e.g., for the (9,~7) tube $\widetilde{n}\in[110, 115]~\mathrm{\mu
m}^{-1}$) and about $50-100~\mathrm{\mu m}^{-1}$ for SWCNTs with
diameters $\sim1.5-2$~nm (e.g., for the (28,~0) tube
$\widetilde{n}\in[50, 55]~\mathrm{\mu m}^{-1}$). The discrepancies
in values of $\widetilde{n}_\mathrm{c}$ obtained from experimental
data and those estimated using relation~(\ref{3.2}) may be
stipulated by ignoring the collective effects in exciton gas and
effects of dynamical screening of the e-h interaction potential.

It is also worth mentioning, that in the considered range
$\varepsilon_\mathrm{env}\in[1,~1.75]$ of seeming instability of
the single-electron spectrum the unstabilized (calculated without
the described stabilization) binding energies of excitons in the
ground state in different SWCNTs obey relation~(\ref{3.1}) with
$\alpha=1.121$ (see figure 1(c)), the respective binding energies
of excitons in the lowest excited states satisfy~(\ref{3.1}) with
$\alpha=1.838$ (see figure 1(d)).

\section{Summary}
In summary, the spectra of excitons in SWCNTs have been studied
within the effective-mass and long-wave approximations and
elementary potential model on the basis of zero-range potentials
method~\cite{adtish},\cite{tish}. These spectra are highly
influenced by the dielectric environment surrounding a nanotube.
The obtained binding energies $\mathcal{E}$ of excitons in the
ground state and the differences between the ground and first
excited exciton energy levels in nanotubes surrounded by medium
with permittivity $\varepsilon_\mathrm{env}\sim4$ are in good
accordance with the corresponding experimental data
from~\cite{maul} and~\cite{wang}. Also, in the range of
$\varepsilon_\mathrm{env}\in[4,~16]$ the ground-state exciton
binding energies $\mathcal{E}$ obey the relation
from~\cite{pereb}:
$\mathcal{E}\sim\varepsilon^{-\alpha}_\mathrm{env}$, where
$\alpha=1.4$. However, in the ranges of permittivities
$\varepsilon_\mathrm{env}\in[1,~1.75]$ and
$\varepsilon_\mathrm{env}\in[2,~4.5]$ these binding energies
satisfy the mentioned relation with slightly smaller values of
$\alpha$: 1.121 and 1.258, respectively. These results are very
close to those from~\cite{wal}, in which $\alpha=1.2$ was obtained
for the whole interval $\varepsilon_\mathrm{env}\in[1,~4]$ using a
model, in which SWCNT was represented as a dielectric cylinder
with some internal permittivity surrounded by a medium with
another dielectric constant. In contrast to our model the nature
of high internal nanotube permittivity in the region of low
environmental permittivities is not explained in~\cite{wal} (there
is only an estimate). However, the conclusion about the important
role of $\varepsilon_\mathrm{int}$ in the exciton parameters
calculation for this region of $\varepsilon_\mathrm{env}$ is made
in~\cite{wal} and this also explains the discrepancy in the result
on $\alpha$ with that obtained in~\cite{pereb} using only
$\varepsilon_\mathrm{env}$.

In the range $\varepsilon_\mathrm{env}\in[1,~1.75]$ the
ground-state exciton binding energies $\mathcal{E}$ exceed the
corresponding energy gaps. This leads to instability of the
nanotube single-electron states with respect to formation of
exctions, but due to their high polarizability in the external
electric field the incipient excitons induce the additional
screening effect, which returns the ground-state exciton binding
energy into the respective energy gap and thus stabilize the
nanotube single-electron spectrum. Because of the transformation
of some single-electron states into excitons the edges of the
forbidden band move apart, and this results in the enhancement
(blueshift) of the lowest optical transition energy $E_{11}$ like
in experiments~\cite{kiow},~\cite{ohno}. The corresponding
estimates for $\varepsilon_\mathrm{env}=1$ are in satisfactory
agreement with results of~\cite{kiow}.

Finally, the present work is initially based on the special
version of independent particle theory, namely, the zero-range
potentials method modelling the self-consistent periodic potential
in a nanotube by the system of universal Fermi pseudo-potentials
located at the carbon atoms positions. This method has exhibited
for determination of band structure and optical spectra of carbon
nanotubes very good accordance with the corresponding extended
LCAO calculations and experimental data. However, for explanation
of the stability of SWCNTs band spectra in vacuum and
low-permittivity media with respect to the exciton formation we
had to explicitly refer to many-particle effects, which in
parallel with stabilization resulted in a slight broadening of the
band gap. Note, that this treatment though formally being
alternative to the quasiparticle approach (as in~\cite{spat}) and
band gap renormalization formalism (as in~\cite{ando2}) does not
principally contradict to them and apparently well agrees with the
experimentally defined relative energy parameters of excitons.

\section*{Acknowledgments}
The author is grateful to Prof. Adamyan V.M. for useful and
interesting discussions.


\begin{thebibliography}{99}
\bibitem{fluor}M.J. O'Connell  {\it et al.}, \textit{Science} {\bf 297}, 593 (2002).
\bibitem{bachilo}S.M. Bachilo, M.S. Strano, C. Kittrell, R.H. Hauge, R.E. Smalley, and R.B. Weisman, \textit{Science} {\bf
298}, 2361 (2002).
\bibitem{bachilo2}R.B. Weisman and S.M. Bachilo, \textit{Nano Lett.} {\bf 3}, 1235 (2003).
\bibitem{maul}J. Maultzsch, R. Pomraenke, S. Reich, E. Chang, D. Prezzi, A. Ruini, E. Molinari, M.S. Strano, C. Thomsen, and C.
Lienau, \textit{Phys. Rev.} B {\bf 72}, 241402(R) (2005).

J. Maultzsch, R. Pomraenke, S. Reich, E. Chang, D. Prezzi, A.
Ruini, E. Molinari, M.S. Strano, C. Thomsen, and C. Lienau,
\textit{Phys. Rev.} B {\bf 74}, 169901(E) (2006).
\bibitem{wang}F. Wang, G. Dukovic, L.E. Brus, and T.F. Heinz, \textit{Science} {\bf
308}, 838 (2005).
\bibitem{ando}T. Ando, \textit{J. Phys. Soc. Japan} {\bf 66}, 1066 (1997).
\bibitem{spat}C.D. Spataru, S. Ismail-Beigi, L.X. Benedict, and S.G. Louie, \textit{Appl. Phys.} A {\bf 78}, 1129 (2004).
\bibitem{cap}R.B. Capaz, C.D. Spataru, S. Ismail-Beigi, and S.G. Louie, \textit{Phys. Rev.} B {\bf 74}, 121401(R) (2006).
\bibitem{jiang}J. Jiang, R. Saito, Ge.G. Samsonidze, A. Jorio, S.G. Chou,
G. Dresselhaus, and M.S. Dresselhaus, \textit{Phys. Rev.} B {\bf
75}, 035407 (2007).
\bibitem{wal}A.G. Walsh, A.N. Vamivakas, Y. Yin, S.B. Cronin, M.S. \"{U}nl\"{u}, B.B. Goldberg, and A.K. Swan, \textit{Physica} E {\bf 40} 2375
(2008).
\bibitem{ando2}T. Ando, \textit{J. Phys. Soc. Japan} {\bf 79} 024706 (2010).
\bibitem{jdd}A. Jorio, G. Dresselhaus, and M.S. Dresselhaus (editors), {\it Carbon Nanotubes. Advanced Topics in the Synthesis, Structure, Properties and Applications} (Springer-Verlag, Berlin, Heidelberg, 2008).
\bibitem{pereb}V. Perebeinos, J. Tersoff, and Ph. Avouris, \textit{Phys. Rev. Lett.} {\bf 92}, 257402 (2004).
\bibitem{wang2}Z. Wang, H. Pedrosa, T. Krauss, and L. Rothberg,  \textit{Phys. Rev. Lett.} {\bf 96}, 047403 (2006).
\bibitem{kiow}O. Kiowski, S. Lebedkin, F. Hennrich, S. Malik, H. R\"{o}sner, K. Arnold, C. S\"{u}rgers, and M.M. Kappes, \textit{Phys. Rev.} B {\bf
75}, 075421 (2007).
\bibitem{ohno}Y. Ohno, S. Iwasaki, Y. Murakami, S. Kishimoto, S. Maruyama, and T. Mizutani, \textit{Phys. Rev.} B {\bf
73}, 235427 (2006).
\bibitem{adsmyr}V.M. Adamyan and O.A. Smyrnov, \textit{J. Phys. A: Math. Theor.} {\bf
40}, 10519 (2007).
\bibitem{jpcs}V.M. Adamyan, O.A. Smyrnov, and S.V. Tishchenko, \textit{J. Phys.: Conf. Ser.} {\bf
129}, 012012 (2008).
\bibitem{adtish}V. Adamyan and S. Tishchenko, \textit{J. Phys.: Condens. Matter} {\bf 19}, 186206 (2007).
\bibitem{tish}S.V. Tishchenko, \textit{Low Temp. Phys.} {\bf 32}, 953 (2006).
\bibitem{aghh}S. Albeverio, F. Gesztesy, R. H\o egh-Krohn, and H. Holden, {\it Solvable Models in Quantum Mechanics. Texts and Monographs
in Physics} (Springer, New York, 1988).
\bibitem{demk}Yu.N. Demkov and V.N. Ostrovskii, {\it Zero-Range Potentials and Their Applications in Atomic Physics} (Plenum, New York,
1988).
\bibitem{banyai}L. Banyai, I. Galbraith, C. Ell, and H. Haug, \textit{Phys. Rev.} B {\bf
36}, 6099 (1987).
\end{thebibliography}
\end{document}